\documentclass[pre,nofootinbib,twocolumn,floatfix,showpacs]{revtex4}
\usepackage{graphicx}
\usepackage{amsmath}
\bibpunct[, ]{[}{]}{,}{a}{,}{,}
\ifnum\lefthyphenmin<2\lefthyphenmin=2\fi
\ifnum\righthyphenmin<2\righthyphenmin=2\fi

\begin{document}

\title{Pseudoknots in a Homopolymer}

\author{A. Kabak\c c\i o\u glu and A. L. Stella$^\dagger$}
\affiliation{INFM - Dipartimento di Fisica, Universit\`a di Padova, I-35131 Padova, Italy\\$^\dagger$ Sezione INFN, Universit\`a di Padova, I-35131 Padova, Italy}

\date{\today }

\begin{abstract}
After a discussion of the definition and number of pseudoknots,
we reconsider the self-attracting homopolymer paying
particular attention to the scaling of
the pseudoknot number ($N_{pk}$) at different temperature
regimes in two and three dimensions.
We find that, although the total number of pseudoknots is extensive
at all temperatures, the number of those forming between
the two halves of the chain diverges logarithmically
at (both dimensions) and below (2d only) the $\theta$-temparature.
We later introduce a simple model that emphasizes the role of
pseudoknot formation during collapse. The resulting phase diagram involves
swollen, branched and collapsed
homopolymer phases with transitions between each pair.
\end{abstract}
\pacs{36.20.Ey, 87.15.Aa, 05.40.-a}
\maketitle
\section{Introduction}


Fueled mainly by the advance in RNA structure determination techniques,
recently there has been growing interest in understanding and predicting
formation of pseudoknots in RNAs.
A pseudoknot (PK) is not a true knot
in the conventional sense. It is a simpler construct generated
by a polymer's self-contacts (see the definition below), therefore
is encountered more frequently.
Unlike true knots which are problematic for the DNA and occur
occasionally in shorter biomolecules, PKs {\it are} the tertiary structure
of the folded RNAs. They are known to exist in almost all RNA classes
including transfer, messanger, ribosomal, viral, catalytic and self-splicing
RNAs (see reviews \cite{pk-reviews}). A recent analysis found that they account
for up to 30\% of the bound base pairs in G+C rich RNA sequences \cite{isambert}.
In addition to stabilizing the
fold, PKs are believed to assume functional roles, such as mediating
the binding of the proteins they encode \cite{McPheeters}, labeling
functionally important positions on the coding regions of the mRNA sequence
\cite{Du,Wills}, mediating frameshifting \cite{farabaugh}, etc.

Being a more elementary topological formation than knots, PKs are
relatively amenable to numerical investigation. Nevertheless,
most of the earlier computational tools and recent 
theoretical work on RNA structure prediction take into
consideration only those configurations without PKs
\cite{Nussinov,Zuker,bundschuh-hwa}. This is mostly because
ignoring PKs results in a 'nested' set of equations and, as a
consequence, allows efficient dynamic programming techniques.
The drawback is that their success is limited to secondary structure
prediction only.
And even then, with limited accuracy due, partially, to a
necessary reorganization of the secondary structure contacts to
accomodate the PKs. 
More recently there appeared computational \cite{Rivas,Uemura} and
theoretical \cite{pilsbury,baiesi-stella,dill,leoni}
studies that include PKs into RNA structure prediction algorithms.
Pilsbury {\it et.al.}
suggest a diagrammatic expansion to perturbatively take into account
the PKs \cite{pilsbury}. 
A recent study by Baiesi {\it et.al.} \cite{baiesi-stella} takes a
more physical look on RNA denaturation.
They model the RNA as a homopolymer traversing a two-tolerant walk
on the FCC lattice
and consider walks both with and without PKs \cite{Mg}.
They conclude that 
the sharp second-order denaturation transition observed when PKs are
allowed gives way to a smooth crossover upon their exclusion. This
result emphasizes the thermodynamic relevance of PKs.
Lucas {\it et.al.} \cite{dill} consider
lattice homopolymers again to argue that the denaturation
transition between the pseudoknotted state and the open state
is continuous. Another two-tolerant trail model with pseudoknots and
with a native state consisting of a single hairpin is argued to
denaturate through a first-order transition \cite{leoni}.
Studying the interplay between the PKs and
the transition thermodynamics within the homopolymer context
is a prerequisite for a deeper understanding of the physics of
RNA pseudoknots.

Folding experiments on RNAs suggest
that it is physically more appropriate to attribute
a different binding energy to the contacts that form a PK \cite{tinoco}.
This energy can be tuned by changing the Mg$^{+2}$ concentration in
the solvent.
Unless this energy is prohibitively high (as in the case of very low
Mg$^{+2}$ concentration), calculating the Boltzmann weights
necessitates identifying PKs for an arbitrary
configuration. As we shall see below, this task, though may be
easier in native RNA configurations, is nontrivial for an
arbitrary polymer.

Accordingly, our goal throughout this paper will be
to explore the thermodynamic role assumed by the PKs
in the well-known context of homopolymer collapse by:\\
\ \ 1. providing an analytic definition for the PK number (Section II),\\
\ \ 2. investigating the scaling properties of the PK number
in various regimes of the homopolymer collapse (Section III),\\
\ \ 3. generalizing the Hamiltonian for the self-attracting homopolymer to
include an arbitrary penalty for PKs
and obtaining the corresponding phase diagram (Section IV).

We hope that our results will provide a better understanding to the
nature of the PKs and a new perspective to the
homopolymer collapse transition by locating it in a more general framework
that also includes the branched polymers and their collapse transition.

\section{Counting pseudoknots}\label{defs}
An arbitrary configuration of a polymer chain can be encoded as
a contact map \cite{vendruscolo} which is a binary symmetric matrix. The
contact map, in turn, can be represented graphically by an 'arc diagram'
as follows:
Imagine stretching the polymer
into a horizontal straight line by pulling from the two ends.  Next,
connect each pair of monomers that were originally in contact
by a semicircular arc on the upper half
plane (the diagram is drawn on a
plane even though the polymer may be embedded in arbitrary
dimensions).

\begin{figure}[h!]
\begin{center}
\end{center}
\includegraphics[width=8cm]{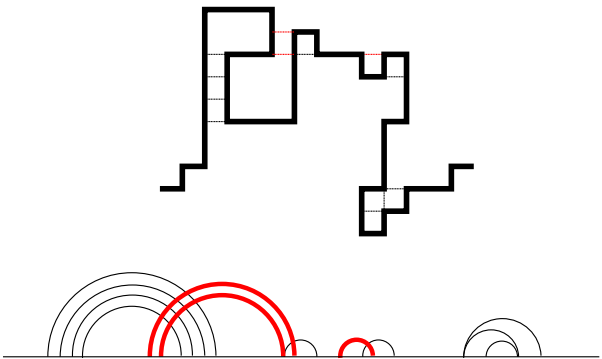}
\caption[]{A sample SAW on a square lattice and its self-contacts
(edges not traversed by the walk that connect two visited nearest-neighbor
lattice sites)
together with the corresponding arc diagram and its non-unique minimal
contact set (shown in bold) to be removed in order to reduce it to a planar
diagram.}
\label{arc-diagram}
\end{figure}
The diagram is said to be {\it planar} if
no two arcs cross each other. This is equivalent to having no PKs.
In the opposite case, a selective treatment of the PKs primarily requires
identification of their number. Since the definition involves crossings in
the arc diagram, one might be inclined to simply count the number of
crossings. However, this does not make physical sense, because an arbitrary
number of crossings may be generated by the addition of a single contact, as
is obvious from Fig.\ref{arc-diagram}. Instead we need a quantity that
reflects the number of contacts that are responsible for the pseudoknots.
Accordingly, we define $N_{pk}$, the
PK number, as the {\it minimum} number of arcs that need to be
removed to reach a planar diagram. The same definition was recently adopted
in another study on RNA pseudoknot prediction \cite{isambert}.
We stress that the choice of this minimal set is in general not unique. 
In Fig.\ref{arc-diagram}, we show a SAW in two
dimensions and the corresponding arc diagram, where one possible choice of
a minimal set of arcs which, when removed, leave a planar diagram is
drawn in bold. Therefore, although one can talk about a unique PK number,
labeling some of the contacts as PK forming contacts requires adopting
an extra arbitrary convention. In this study, we avoid this by resting our
results on the mere knowledge of the number of PKs.

\begin{figure}[h!]
\begin{center}
\end{center}
\includegraphics[width=5cm, angle=0]{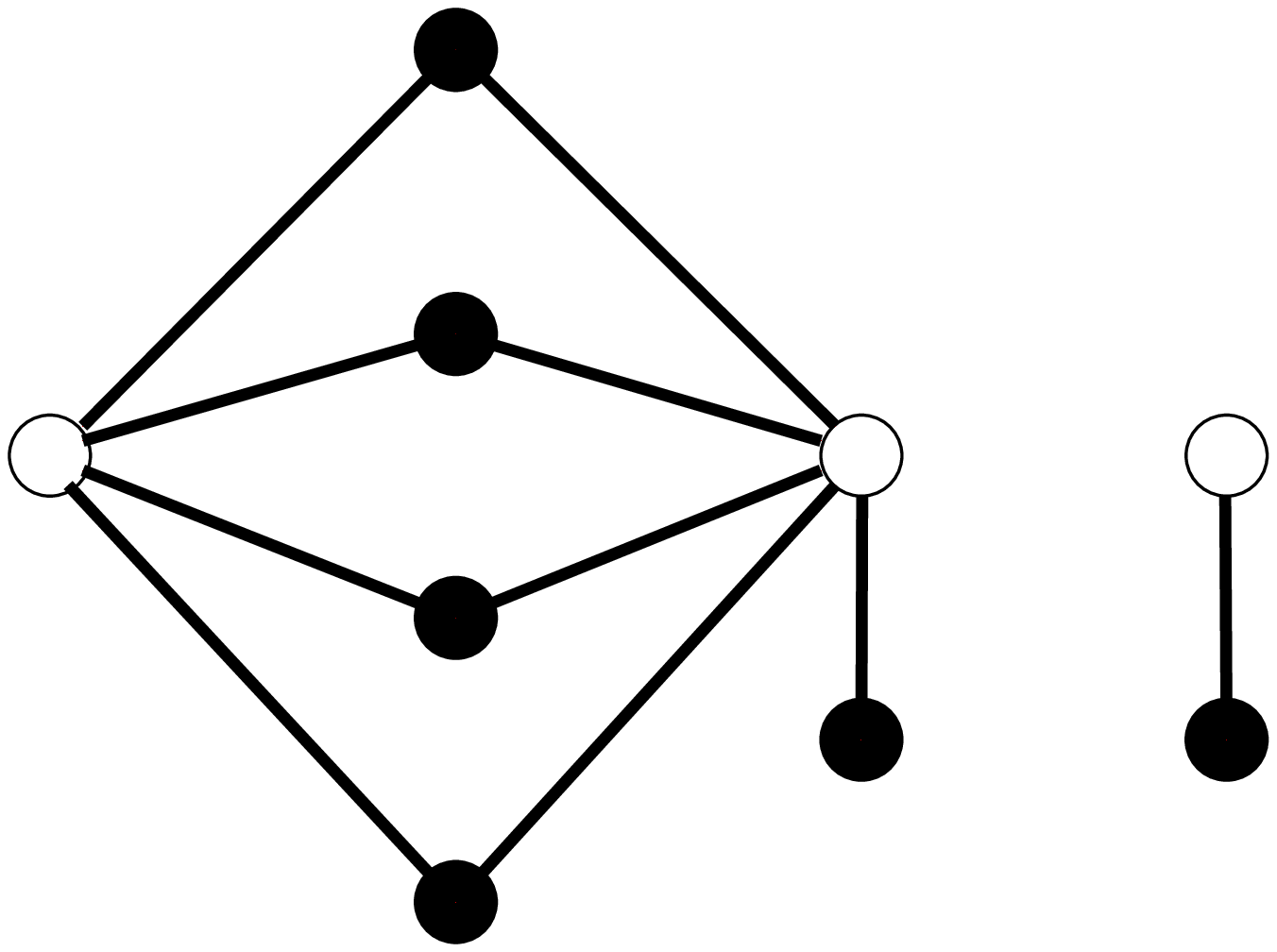}
\caption[]{The resulting graph for the arc diagram in Fig.\ref{arc-diagram}
after the mapping to the vertex-covering problem. Empty vertices correspond
to the solution represented by bold arcs in Fig.\ref{arc-diagram}.}
\label{mapping}
\end{figure}

Our first observation is that calculating $N_{pk}$ exactly
for an arbitrary arc diagram belongs to a class of problems
known as {\it NP-complete}, 
implying that there's no known deterministic polynomial-time algorithm
for calculating $N_{pk}$ \cite{NP-complete}. We prove this by mapping our problem to one of the
six well-known problems in computer science that are shown to be NP-complete,
namely the 'vertex-covering' problem. For a recent review of the vertex-covering
problem in the statistical physics context see \cite{hartmann}. The mapping
is easily established by 
representing each arc by a vertex and drawing edges 
between pairs of vertices corresponding to crossing arcs 
(see Fig.\ref{mapping}).
The vertex-covering
problem on the obtained graph (known as the incompatibility graph) amounts
to finding a minimal set of vertices
which, when labelled, results in labelling at least one end of every edge.
In other words, erasing those vertices alone together with the
edges sprouting from each is sufficient to get rid of all the edges in 
the graph.
Since every edge reflects a crossing and every vertex an arc
in the original arc diagram,
eliminating those arcs corresponding to the minimal vertex set obviously
results in a planar diagram, i.e., the size of this minimal set is equal
to $N_{pk}$.
Since we need to calculate
$N_{pk}$ for many polymer configurations in our study, it is important to
be aware that an exact treatment requires CPU times exponential in the
number of arc-crossings.
The above mapping is
a different and simpler statement of NP-completeness than an earlier
proof that a large class of RNA secondary structure
prediction algorithms based on free energy minimization with pseudoknots
are NP-complete \cite{lyngso}. Also note that when restricted to a certain
subset of polymer configurations (typically generated in an iterative manner),
the problem can be solved in polynomial time. Efficient algorithms on such
restricted configurations have been recently utilized to predict RNA
pseudoknots \cite{Rivas,isambert,pilsbury}.

What is the expected value of $N_{pk}$ for a polymer?
We would first like to provide some
insight for the reader by calculating $N_{pk}$ for a random arc diagram:
Consider $N_c$ randomly chosen distinct pairs of points on a line segment
and connect them by arcs as in Fig.\ref{arc-diagram}b.
The probability that $m$ randomly placed arcs do not intersect (i.e., form
a planar graph) is
\begin{equation}
P_{planar}(m) \sim 1/(m!)\ .
\label{induction}
\end{equation}
Proof by induction:
the $m^{\mbox{th}}$ non-intersecting arc to be placed will
see the polymer divided into $m$ compartments, such that,
if the arc is placed with its legs resting on different compartments,
it has to cross at least one other arc.
Then, the probability that  the $m^{\mbox{th}}$ arc doesn't cross
(given the first $m-1$ arcs) is roughly $1/m$, i.e., the probability that
the two points fall into the same bin. Eq.(\ref{induction}) follows
by induction.

Then, such a subset of $m$ planar arcs will be found
among available distinct subsets when
\begin{equation}
\binom{N_c}{m} \ge 1 / P_{planar}(m)\ .
\end{equation}
$m$ is maximized when the two sides are equal.
Applying Sterling's approximation on both sides
leads to $m_{max} \propto \sqrt{N_c}$. In other words,
the typical number of pseudoknots for a random arc diagram
approaches the number of contacts in the thermodynamic limit
($N_c \rightarrow \infty$) as
\[
(N_c - \langle N_{pk} \rangle) / N_c \propto 1/\sqrt{N_c} .
\]

Yet, true polymers and lattice walks do not come with random arc diagrams.
There is a considerable correlation among the contacts due to the existence of
an underlying chain and the effective repulsion resulting through self-avoidance,
both of which favor contacts between monomers that are closer along the chain.
This tendency is reflected in the loop length distribution
\cite{vanderzande},
\begin{equation}
\label{distribution}
	P(l) \propto l^{-c} ,
\end{equation}
where $c=d/2$ (random-walk), $c=d\nu - \sigma_4=2.68, 2.22$ (SAW in 2d, 3d),
and $\sigma_4$ the critical dimension associated with a 4-leg vertex as in
the polymer network theory of Duplantier \cite{Duplantier}.
Correspondingly, one expects $N_{pk}$ for the real chains
to be less than the above 'random graph' value.
The combinatorial argument presented above picks
random pairs of points along the chain with equal probability
irrespective of the distance in between ($c=0$).
Unfortunately, it does not generalize easily to $c > 0$.
However, our numerical analysis on random graphs with arbitrary $c$
suggests
\[
\langle N_c - N_{pk} \rangle \propto N_c^{q(c)} ,
\]
with $q(c)$ increasing almost linearly from $q(0)=1/2$ and saturating at
$q(\sim 2) = 1$.
The interesting statistical properties of the incompatibilty graph with
the distribution in Eq.(\ref{distribution}) will be reported elsewhere.
A further reduction
is expected in two dimensions due to additional constraints imposed by
the impenetrability of encircled regions: The fact that each polymer contact
divides the plane to two disconnected regions translates to having a bipartite
incompatibility graph. As a consequence, $N_{pk} \le N_c/2$ (since the
arc-diagram turns out to be planar when arcs are allowed on both half-planes
instead of one).
Unlike the general case, the vertex-cover problem on
bipartite graphs is solvable in polynomial time \cite{graph-theory}.

One can obtain a lower-bound on $N_{pk}$ from Kesten's Pattern Theorem
\cite{madras}, i.e. by noting that a local pseudoknotted pattern (e.g.
the shortest S-shaped walk on a square lattice) has a finite probability
of occurence in an infinite chain. Therefore 
\begin{equation}
	\langle N_{pk} \rangle >  a N
\end{equation}
for a walk of $N$ steps and for some $a > 0$. 
Extensivity of $\langle N_{pk} \rangle $ for a homopolymer
supports the observation that their exclusion may
have manifestations on the nature of the transition in the thermodynamic
limit \cite{baiesi-stella} and
that penalizing pseudoknot formation can change
the low-temperature phase from collapsed to a branched polymer
(Section.\ref{sec-phase-diagram}).

\section{Pseudoknot Numerics}\label{numerics}

In this part of our analysis we look at the ordinary homopolymer
collapse, where the energy is not sensitive to the pseudoknot formation.
We obtained statistics numerically for self-avoiding chains
of typical size $N=300$, even though
we checked for size independence of our results occasionally by going up
to $N=800$. All our results were obtained by using an improved
version of the PERM algorithm developed by Grassberger {\it et.al.}
\cite{grassberger}.
Although with PERM it is typical to simulate much longer homopolymers,
our statistics were mainly limited by the fact that $N_{pk}$ for each
configuration has to be calculated from scratch, unlike, e.g., the number of
contacts which is updated incrementally at each step of the walk. We present
results for square and cubic lattices in two and three dimensions, respectively.

\begin{figure}[h!]
\begin{center}
\end{center}
\includegraphics[width=8cm, angle=-90]{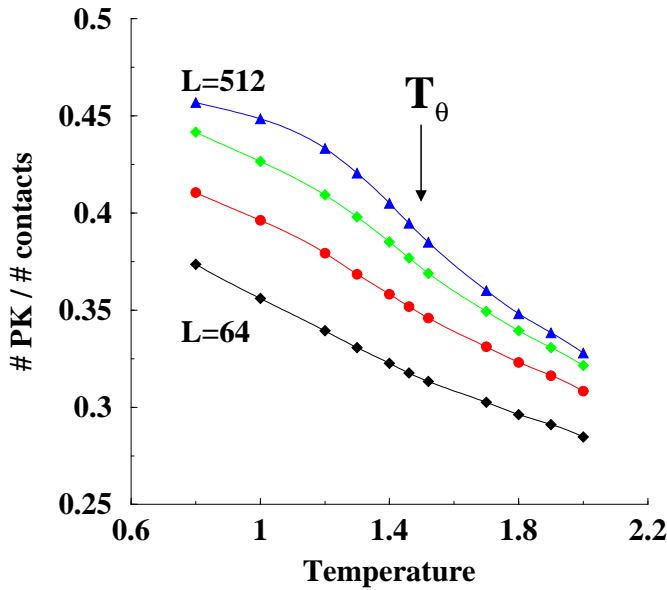}
\caption[]{Fraction of pseudocontacts in a SAW as a function of temperature
for different walk lengths in two dimensions.}
\label{prefactor}
\end{figure}

It is possible to calculate $N_{pk}$ using an exact back-tracking
search algorithm which is straightforward to implement, but
requires a runtime exponential in system size. 
Since we want to obtain statistics for reasonably long chains, it is not
feasible to use such an exact method.
Instead, we calculate $N_{pk}$ approximately by means of a greedy algorithm
which at each step eliminates (one of) the maximally crossing arc(s).
The choice is made randomly when they are more than one.
Due to the stochastic nature of the algorithm and the fact that the optimal
selection may involve eliminating a less than maximally crossing arc,
this greedy algorithm provides an upper bound to $N_{pk}$.
For details of various exact algorithms and the above greedy algorithm we
refer the reader to \cite{hartmann}.

For comparison, we also implemented an exact calculation of $N_{pk}$ on
the typical diagrams we encountered.
We found that the average deviation  from the exact value of
the upper bound on $N_{pk}$ obtained by the greedy algorithm,
although increasing with growing chain length, approaches
a constant fraction around $1.5\%$ of the exact $N_{pk}$. Therefore, we
are confident that our conclusions concerning the scaling behavior are not
effected by the approximate algorithm we adopted.

At all temperature regimes, we find that $\langle N_{pk}\rangle$
is a fraction of the total number of contacts with a
temperature-dependent proportionality constant, $a(T)$. Although a
Fermi-function-like limiting behavior is evident from Fig.\ref{prefactor},
the precise form of $a(T)$ in the thermodynamic limit requires a more
elaborate analysis which we will not attempt here.

$a(T)$ reveals the leading behavior in $\langle N_{pk}\rangle$, however theory of
critical phenomena has taught us that the non-trivial behavior of many
systems is reflected in the non-analytic contribution to the extensive
quantities. Recently, Orlandini {\it et.al.} \cite{orlandini} considered
the scaling of contacts
formed between the two halves (referred from here on as A and B)
of a SAW at the $\theta$-point as a direct and
precise way of measuring the crossover exponent, $\phi_\theta$. 
The number of such contacts scales as
\begin{equation}
\label{AB_scaling}
\langle N^{AB}_{c}(T_\theta) \rangle \propto  N^{\phi_\theta} ,
\end{equation}
where $\phi_\theta=3/7$ in two dimensions, as can be
shown analytically by using a recent extension of Saleur-Duplantier results
for polymer criticality \cite{Aizenmann,Duplantier}.
The value of the crossover
exponent is typically difficult to confirm by numerics, because, as a rule,
it has to be extracted from subleading singular terms, when considring the
set of all contacts along the chain \cite{note-subdominant}.
Focusing on the contacts
between the two halves strongly filters out the dominant analytical
contribution of the local contacts along the chain, thus surfacing the
otherwise concealed non-analyticity. AB-contact statistics has been
fruitful in a variety of polymer models \cite{orlandini,carlon}.

We use the same method here to pinpoint the singularity of $\langle N_{pk}\rangle$, which
at the leading term scales identical to the contact energy (extensivity of
$\langle N_{pk}\rangle$ due to Kesten's Pattern Theorem).
More precisely, we start by identifying all A-B contacts, $N^{AB}_c$,
as above, and then calculate the number
of A-B pseudoknots, $N^{AB}_{pk}$, by eliminating the crossing arcs
in the arc diagram corresponding to A-B contacts only.
Since $\langle N_{pk}\rangle$ on the whole chain is a fraction of the contacts,
$\langle N_c\rangle$, at all temperatures,
one may expect that $\langle N^{AB}_{pk}\rangle$ should also scale identical
to $\langle N^{AB}_c\rangle$ at all temparatures.
In contrast, we find a qualitatively different
scaling of the A-B pseudoknot number.\\

Let us consider each temperature regime separately:\\

\begin{figure}[h!]
\begin{center}
\end{center}
\includegraphics[width=8cm, angle=0]{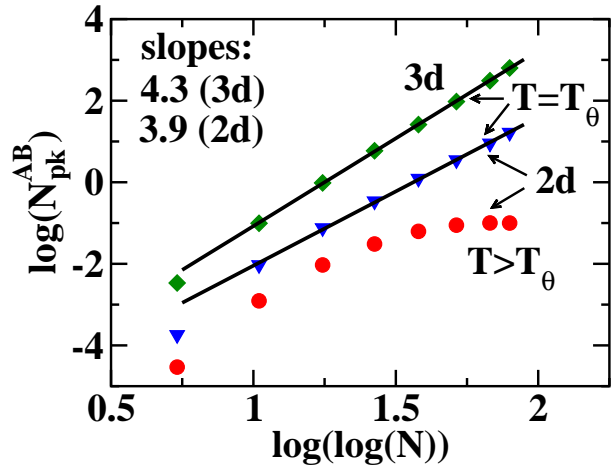}
\caption[]{Scaling of the A-B pseudoknot number for
$T>T_\theta$ (2d), $T=T_\theta$ (2d) and $T=T_\theta$ (3d).
Estimated asymptotic slopes are $v_\theta \sim 0$,$3.9$ and $4.3$, respectively.
Maximum walk size was 800 steps in both dimensions.}
\label{log-scaling}
\end{figure}

Above the $\theta$-point ($T > T_\theta$), no surprises are expected: 
$\langle N^{AB}_c\rangle$ already saturates
to a constant in both two and three dimensions.
Since $N^{AB}_{pk} < N^{AB}_c$, this leaves no alternative
to $\langle N^{AB}_{pk}\rangle$ but to stay finite as well. This is confirmed by our
numerics in Fig.\ref{log-scaling}.\\

At the $\theta$-point, the PERM algorithm is the most efficient. Therefore
we expect our results to be most accurate in this region. Recall that
$\langle N^{AB}_c \rangle \propto N^{3/7}$ for $T = T_\theta$, which can be obtained
analytically and verified numerically to high accuracy.
Our numerical results for $\langle N^{AB}_{pk}\rangle$ on the other hand indicate
a logarithmic divergence of the form
\begin{equation}
\label{log-behavior}
\langle N^{AB}_{pk}(T_\theta) \rangle \propto  (\log{N})^{v_\theta} ,
\end{equation}
where to our best estimate, $v_\theta \sim 3.9$
in two dimensions (Fig.\ref{log-scaling}). In three dimensions,
the logarithmic behavior survives (only at $T=T_\theta$), albeit
with a different exponent $v_\theta \sim 4.3$.

In retrospect, one could argue a priori that $\langle N^{AB}_{pk}\rangle$ and
$\langle N^{AB}_c\rangle$ should have qualitatively different scaling properties
at the $\theta$-temperature:\\

\begin{figure}[h!]
\begin{center}
\end{center}
\includegraphics[width=8cm, angle=0]{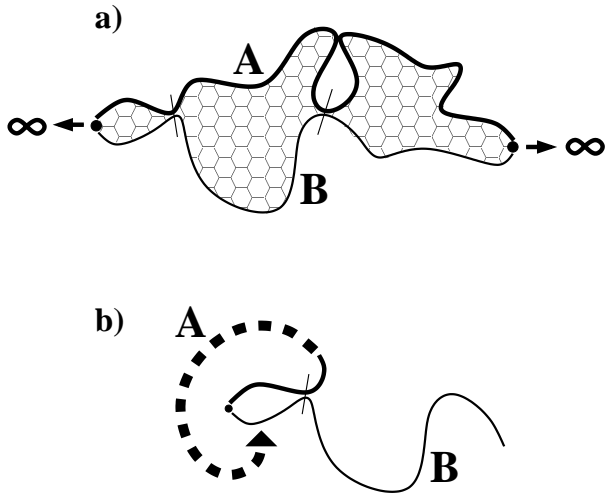}
\caption[]{The curves A(thick) and B(thin) represent the two halves of the
full hull of a percolating cluster at the percolation transition.
The number of AB-contacts (indicated by the cutting line
segments) scales as $N^{3/7}$ in the thermodynamic limit when the
two diametrically opposite points on the hull bordering A and B
are taken to infinity (a). An AB-pseudoknot can not be
described in this setting once the thermodynamic limit is taken (b).}
\label{pk_and_hull}
\end{figure}

We recall that the easiest way to obtain $\phi_\theta$ is to use
the correspondence between a ring polymer at the $\theta$-point
and the full hull of a
percolating cluster at the percolation transition in two dimensions
\cite{orlandini}. In this picture, the AB-contacts correspond to the
the 'red' contacts between the two halves that lie between
two diametrically opposite points of the hull (which are at the
opposite infinities, see Fig.\ref{pk_and_hull}a) \cite{carlon}.

The key observation is that, an AB-pseudoknot cannot be formed
'locally' between the two halves, because one of the two halves
of the chain should wrap around the midpoint to
make a pseudoknot-forming A-B contact, as shown in Fig.\ref{pk_and_hull}b.
This imposes a rather stringent
condition on the A-B pseudoknot formation. The numerical result of
Eq.(\ref{log-behavior}) nevertheless indicates a logarithmic divergence
for $\langle N^{AB}_{pk}\rangle$. The scenario depicted in
Fig.\ref{pk_and_hull}b suggests a likely connection to the statistics of
the homopolymer winding angle, $\omega$, for which
\begin{equation}
\label{log-behavior-winding}
\langle \omega^2 \rangle \propto \log{N}
\end{equation}
in the swollen phase and at the $\theta$-point in two dimensions
\cite{kardar-drossel}. Yet, this would have the interesting implication
that the similar log-scaling observed in three dimensions has a different
origin. These possibilities will be investigated in the future.
\\

Below the $\theta$-point and in two dimensions,
the logarithmic growth of $\langle N^{AB}_{pk}\rangle$ appears to persist.
Although the numerics in this region is not as reliable, the
lack of a power-law behavior similar to that for $\langle N_{pk}\rangle$
is not surprising due to the above geometric considerations.
Note that the scaling of $\langle N^{AB}_c\rangle$ in this regime is still
power-law \cite{dense_phi}.
In three dimensions and for $T < T^{3d}_\theta$, preliminary calculations
suggest $\langle N^{AB}_{c} \rangle \propto N$ and a deviation
from the logarithmic behavior in $\langle N^{AB}_{pk} \rangle$. This is
probably due to the fully A-B co-penetrated configurations of the compact
polymer in 3d. Unlike in 2d, the A-B boundary fills the volume.

\section{Pseudoknot-sensitive homopolymer collapse}\label{sec-phase-diagram}

\begin{figure}[h!]
\begin{center}
\end{center}
\includegraphics[width=8cm, angle=0]{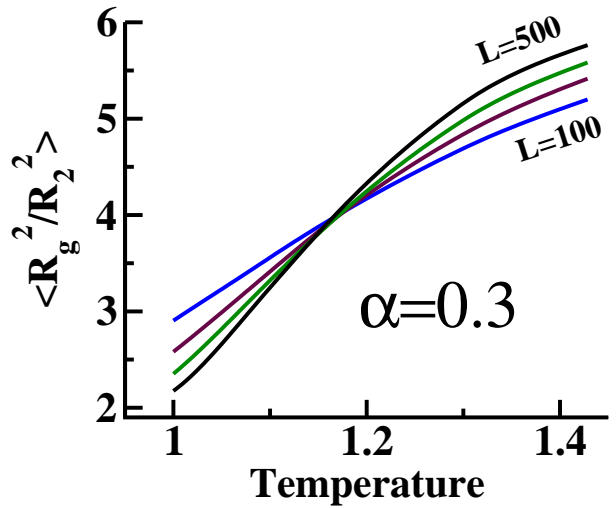}
\caption[]{Transition temperature for each value of $\alpha$ is located
as the crossing point of $\langle R_e \rangle / \langle R_g \rangle$ curves
as a function of temperature for different polymer lenghts. The curves is
an interpolation between the data points for two dimensions and $\alpha=0.3$.}
\label{universal-ratio}
\end{figure}

The pseudoknot formation is essential for the collapse of a polymer to
a compact structure. This is most easily seen by comparing the
radius of gyration, $R_g$, for a self-attracting homopolymer with that
for which the pseudoknot forming contacts are excluded from the
energy calculation. Consider the following generalized Hamiltonian for
a self-avoiding lattice walk:
\begin{equation}
\label{Hamiltonian}
H = \sum_{ij}{\epsilon_c\,\Delta(i,j) - \alpha\,\epsilon_c\,N_{pk}}
\end{equation}
with $\Delta(i,j) = 1$ for pairs of occupied nearest-neighbor lattice sites
not consecutive along the walk and $0$ else. $\alpha = 0,\infty$ correspond
respectively to the usual
self-attracting homopolymer with and without pseudoknots. $\alpha=1$ is
the case when the homopolymer energy is given by the number of contacts
in the maximal planar 'sub'-arc-diagram.
In Eq.(\ref{Hamiltonian}), we deliberately avoided writing down $N_{pk}$ as
a sum over contacts, since only the number $N_{pk}$ and not the identity
of the PK-forming contacts is well-defined.

We located the transition point for different values of
$\alpha$ by plotting
$\langle R_e \rangle / \langle R_g \rangle$, the ratio of the averaged
end-to-end distance and the radius of gyration {\it vs.} temperature.
This universal ratio as a function of temperature should converge to a 
step function in the thermodynamic limit ($N\rightarrow\infty$) with
a universal intermediate value at the $\theta$-point \cite{vanderzande,privman}
The crossing point of the curves in Fig.\ref{universal-ratio}
for different chain sizes is an efficient way of locating the transition
temperature and the critical universal ratio at the transition
\cite{baiesi-stella}. The resulting phase-diagram is given in
Fig.\ref{phase-diagram}.

\subsection{Phase Diagram}

\begin{figure}[h]
\begin{center}
\end{center}
\includegraphics[width=9cm]{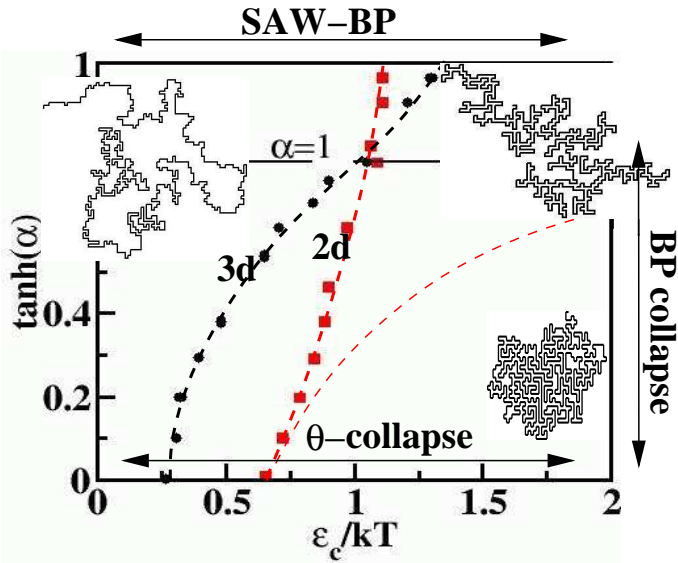}
\caption[]{The phase-diagram for the Hamiltonian in Eq.\ref{Hamiltonian}.
The lower horizontal axis corresponds to the usual homopolymer with
self-attraction. SAW-BP transition line were obtained from the numerical
data by the universal-ratio crossing method described in the text. The
terminal point on the lower axis of the transition curves correspond to
the $\theta$-point for that dimension. Dashed curve is the expected
BP-CP boundary with a limiting (T$\rightarrow$0) value of $(1-\alpha)/T=0.69$.}
\label{phase-diagram}
\end{figure}

Typical low-temperature configurations in the limit $\alpha \to \infty$
are double-stranded branched structures as shown in the inset of
Fig.\ref{phase-diagram}.
In fact, one can show that the ground-state configurations in
this limit are the nnn-avoiding lattice-trees. For a proof
in two dimensions, it is sufficient to note that the number of energetically
favorable contacts (size of the maximal planar sub-diagram) is maximized
when the two ends of the walk meet at nearest-neighbor sites to
form a fully deflated self-avoiding ring. Thus at zero temperature, the
Hamiltonian in Eq.(\ref{Hamiltonian}) with $\alpha = 1$ is equivalent to
the Leibler-Singh-Fisher (LSF) model \cite{fisher} of planar vesicles
with negative area fugacity. LSF model with negative pressure is established
to have a BP low-temperature phase.
The corresponding BP lives on the dual-lattice points inside the ring
and the self-avoidence of the ring translates to next-nearest-neighbor-avoiding
branches in the dual lattice. 
Consistently, we numerically obtain for the
radius-of-gyration exponent, $\nu(T<T_c) \sim 0.62$,
very close to the value of $\nu_{BP} = 0.64$.
SAW$\rightarrow$BP transition was studied earlier in several
lattice polymer models \cite{orlandini-seno,dekeyser-orlandini, stella-bp}.
We note that the Hamiltonian in Eq.(\ref{Hamiltonian}) 
exhibits an SAW-BP transition also in higher dimensions, especially
in three-dimensions which could be relevant to RNA-folding. The
zero-temperature mapping to lattice trees presented above applies to three
dimensions as well, although the dual lattice on which the corresponding
branched polymers live is not simply the shifted simple cubic lattice.

The low-temperature scaling of $\langle R_g \rangle$ along the
line $\alpha = 1$ is still BP-like. Deep in the BP phase,
$ \epsilon^{pk} \equiv (\alpha-1)\,\epsilon_c$ acts as a contact interaction
between the branches (negative $\alpha$ being the attractive regime).
Considering earlier studies on BPs \cite{flavio},
we then expect a second transition line in each dimension between
two low-temperature phases, BP and the collapsed polymer (CP),
as shown in dashed-line for two dimensions only in Fig.\ref{phase-diagram}.
The simplest scenario is that the BP-CP boundary splits from the
SAW-BP boundary
at the $\theta$-point and asymptotically approaches the $\alpha=1$
line such that $\epsilon^{pk}/T = const$. The critical interaction 
for the collapse of lattice-trees has been extensively studied
in the literature \cite{madras-rensburg,flavio} for both dimensions.
We merely speculate this section of the phase diagram, since obtaining
good statistics at temperatures low enough to distinguish the two
phases was not possible.

An interesting feature of the phase diagram in Fig.\ref{phase-diagram}
is the crossing of the phase boundaries corresponding to two and three
dimensions around $\alpha = 1$, reflecting the fact that the
$T_\theta$ for a self-attracting homopolymer ($\alpha = 0$) increases with
increasing dimension, whereas the SAW$\rightarrow$BP transition temperature
for $-\alpha \gg 1$ has the opposite trend!
The transition temperature in each case is determined by
the interplay between the entropy of the coil and the energy of
the collapsed state. For $\alpha = 0$, as we switch from the square
to the cubic lattice, the increased gain in contact energy by collapse
(due to the higher number of nearest-neighbor sites) overshadows the
increased loss of entropy (due to -roughly- the change in the
connectivity constant).
As a result, $k_B\,T_\theta/\epsilon_c$ moves up from
1.5 to 3.5. In the other limit (-$\alpha \gg 1$), the contact energy due to the
partial collapse to a BP is proportional to N and 
{\it independent of dimension}. Yet, the entropy loss due to collapse
still increases with
dimensionality. Then, the collapse to a BP should happen at a lower
temperature with increasing dimension. The two opposing trends cancel
each other around $\alpha=1$.

Also note that, positive but small values of $1-\alpha$ describe a transition
upon reducing the temperature first to a branched-polymer-like state followed by
a PK-mediated collapse. Such collapse (or melting) of RNAs with an
intermediate pronounced with lowered Mg$^{+2}$ concentration
has been experimentally observed \cite{tinoco,treiber,pan} and discussed in \cite{leoni}.

\section{Conclusions}\label{conclusions}

To summarize, we attempted in this paper to provide a mathematical
definition for the {\it number} of pseudoknots, $N_{pk}$, in a polymer chain.
With this definition, we show that counting the number of pseudoknots
is equivalent to
the well-known 'vertex-cover' problem which is NP-complete.
Nevertheless, it is possible to study numerically the statistical
properties of pseudoknots by employing an efficient approximate scheme.
We show that the average total number of pseudoknots is extensive at all
temperatures, however the number of pseudoknots forming between
the two-halves of the chain scales logarithmically with chain size at the
$\theta$-point of a homopolymer in both two and three dimensions, and
also for $T < T_\theta$ in two dimensions. This logarithmic character
is likely be related to the winding-angle statistics in two dimensions.

We also studied the role of pseudoknots in the homopolymer collapse
by considering a Hamiltonian which favors polymer self-contacts but
penalizes pseudoknots. We showed that in the absence of an energetic
preference for pseudoknot formation, the low-temperature phase is
a branched-polymer. When the ratio of two competing energies
satisfies $0 < \alpha < 1$, Hamiltonian (\ref{Hamiltonian}) allows
a transition scenario from a SAW to a collapsed phase with an intermediate
branched polymer regime, where the BP-CP transition is
mediated by pseudoknot formation. The critical properties of the
phase diagram and relevance to RNA folding need further investigation.

\section{Acknowledgements}

A.K. thanks D. Yuret and C.E. Soteros for pointing out, respectively, the
NP-complete character of the problem and the existence of a polynomial-time
solution on bipartite graphs. We acknowledge support from INFM-PA02 and
MIUR-COFIN01.


\begin{thebibliography}{99}
\bibitem{Mg}{There is experimental evidence that these regimes can be
reached by modifying the Mg$^{+2}$ concentration in the solution. E.g., see
V.P. Misra and D.E. Draper, Biopolymers {\bf 48}, 113 (1999).}
\end{thebibliography}
\end{document}